\documentclass[prb,twocolumn,groupedaddress,showpacs]{revtex4}
\usepackage{graphicx}
\begin{document}
\title{The network topology of a potential energy landscape: A static scale-free network}
\author{Jonathan P.~K.~Doye}
\email{jon@clust.ch.cam.ac.uk}
\affiliation{University Chemical Laboratory, Lensfield Road, Cambridge CB2 1EW, United Kingdom}
\date{\today}
\begin{abstract}
Here we analyze the topology of the network formed by the minima and transition states on the
potential energy landscape of small clusters. We find that this network has both a small-world
and scale-free character. 
In contrast to other scale-free networks, where the topology results from the 
dynamics of the network growth, the potential energy landscape is a static entity.
Therefore, a fundamentally different organizing principle underlies this behaviour:
The potential energy landscape is highly heterogeneous with the low-energy minima 
having large basins of attraction and acting as the highly-connected hubs in the network. 
\end{abstract}
\pacs{89.75.Hc,61.46.+w,31.50.-x}
\maketitle

Energy landscapes have been at the forefront of many of the recent theoretical
developments in our understanding of biomolecules,\cite{Brooks01} clusters\cite{WalesMW98,WalesDMMW00} 
and the glass transition.\cite{Debenedetti01}
For example, this research has provided important new insights into how 
proteins fold\cite{Bryngelson95} and the origin of the unusual properties of supercooled liquids,
such as the distinction between ``strong'' and ``fragile'' liquids.\cite{Sastry01,SaikaVoivod01}
There has also been a surge of interest in modelling complex systems as
networks,\cite{Strogatz01} inspired by Watts and Strogatz's discovery that many 
networks behave as ``small worlds''.\cite{Watts98}
Intriguingly, a diverse range of such networks, e.g.\ the world-wide web,\cite{Albert99}
the internet,\cite{Faloutsos99}  scientific collaboration\cite{Newman01a} and citation\cite{Redner98} networks, 
and biochemical networks,\cite{Jeong00,Jeong01} 
also have a ``scale-free'' topology, where 
the distribution of the number of connections to each node, the degree, follows a power law.
This topology results from the dynamics of the network growth in these systems.\cite{Barabasi99}
Here we draw these two strands of research together by applying the techniques of
network analysis to probe the global structure of potential energy landscapes of clusters.

\begin{figure}
\includegraphics[width=8.6cm]{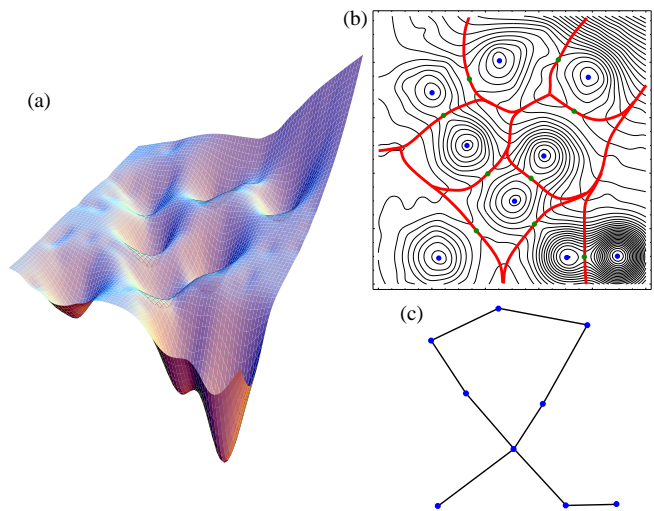}
\caption{
(a) A model two-dimensional energy surface. 
(b) A contour plot of this surface
illustrating the inherent structure partition of the configuration space 
into basins of attraction surrounding minima.  
The basin boundaries are represented by red lines, 
and the minima and transition states by blue and green points, respectively.
(c) The resulting representation of the landscape as a network.
}
\label{fig:PES}
\end{figure}

The potential energy landscape is a multi-dimensional surface 
representing the dependence of the potential energy on the 
positions of all the atoms of the system. 
For a system with many atoms the landscape will have a complex topography with 
higher-dimensional analogues of mountain ranges, valleys and passes.
Although the potential energy landscape determines the system's structure, thermodynamics
and dynamics, the nature of this relationship is complex. 
A particularly successful means of elucidating this relationship
is the inherent structure approach of Stillinger and Weber,\cite{StillW84a} 
in which the landscape is divided into basins of attraction surrounding 
each minimum (See Fig.\ \ref{fig:PES}).
This partition provides a natural way to describe the dynamics of the system, because, except
at high temperature, the system spends most of the time vibrating in the well surrounding
a minimum and only occasionally hops to a different well by passing over a transition state.
The interbasin dynamics can then be represented as a walk on a network whose
nodes correspond to the minima and where edges link those minimum which are directly connected
by a transition state.
Figure \ref{fig:PES} provides an illustration of such an inherent structure network (ISN)
for a two-dimensional energy surface. 

The ISN should provide the starting point for an energy landscape view
of the global dynamics of a system. Indeed, as is increasingly being done,
it is relatively easy to calculate the dynamics from this network using 
a master equation approach.\cite{BerryK95,Cieplak98a,WalesDMMW00} 
However, fundamental questions about the topology of the ISN have received little attention.
By contrast the global topography of energy landscapes has been the 
focus of much research.\cite{Brooks01,WalesMW98,Bryngelson95} 
To give one example, this emphasis has revealed that when
a landscape is like a `funnel'\cite{Bryngelson95} the system is guided 
towards the global minimum, be it the native state of a protein,\cite{Bryngelson95} 
an ordered nanoparticle\cite{Ball96} or a bulk crystal.
However, the topological aspects of this idea remain open despite their importance:
if the average number of steps to reach the global minimum from an
arbitrary starting minimum scales unfavourably with size, 
the location of this structure would become significantly hindered at large size.

\begin{figure}
\includegraphics[width=8.6cm]{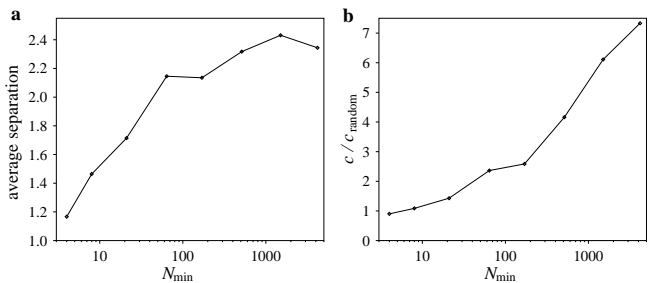}
\caption{
(a) The dependence of the average separation between nodes (in steps) 
on the size of the network, $N_{\rm min}$. 
(b) The size dependence of the clustering coefficient, $c$, compared to that for a random graph,
where $c$ is the fraction of the pairs of nodes with a common neighbour that are also 
connected.\cite{Watts98}
The data points are for Lennard-Jones clusters with from 7 to 14 atoms.
}
\label{fig:SW}
\end{figure}

To characterize the topology of the ISN 
we study small clusters for which we are able to locate nearly all 
the minima and transition states on the potential energy landscape.\cite{WalesDMMW00,Tsai93a}
The atoms of the cluster interact with a Lennard-Jones potential, which provides
a reasonable description for rare gas clusters. 
The numbers of minima and transition states are expected to increase roughly as 
$N_{\rm min}\approx e^{\alpha N}$ and 
$N_{\rm ts}\approx N e^{\alpha N}$, respectively,\cite{Still99,Doye02a} 
where $N$ is the number of atoms in the cluster.
Therefore, the largest network that we are able to consider is for a 14-atom cluster 
for which we have located 4196 minima and $87\,219$ transition states

Small-world networks have characteristics typical of both random graphs and lattices.
The average separation between nodes scales logarithmically with network size,
while the network is highly clustered, i.e.\ any two neighbours of a node 
are also likely to be connected. 
From Fig.\ \ref{fig:SW} it is clear that the ISNs for the clusters
show both these features and so are small worlds. 
The clustering is unsurprising given that the connections between basins on a 
potential energy landscape are based on adjacency in configuration space,\cite{cluster} 
but to interpret Fig.\ \ref{fig:SW}(a) properly we must take into account the increase in both the 
dimension of configuration space and the average degree, $\langle k\rangle$, 
as the size of the network increases. 

For example, for a hypercubic lattice 
with a constant number of lattice points, $L$, along each edge
and dimension $3N$, 
the number of lattice points, $N_{\rm latt}$, 
scales exponentially with $N$ and 
the average number of steps between any two lattice points is 
$3N(L+1)/3=(L+1)\log N_{\rm latt}/ 3 \log L$.
By contrast, if the network behaves as a random graph, 
the average separation should scale as $\log N_{\rm min}/\log \langle k\rangle\propto 
\log N_{\rm latt}/\log(\log N_{\rm latt})$ 
because $\langle k\rangle\propto N_{\rm ts}/N_{\rm min}\propto N$. 
The sublogarithmic scaling suggested by 
Fig.\ \ref{fig:SW}(a) points to the latter scenario.
This result is somewhat surprising.
In Watts and Strogatz's small-world model the random-graph character results from the 
introduction of random links into the lattice, which can potentially connect up distant nodes, 
but there is no obvious equivalent of the random links on the potential energy landscapes.

\begin{figure}
\includegraphics[width=8.6cm]{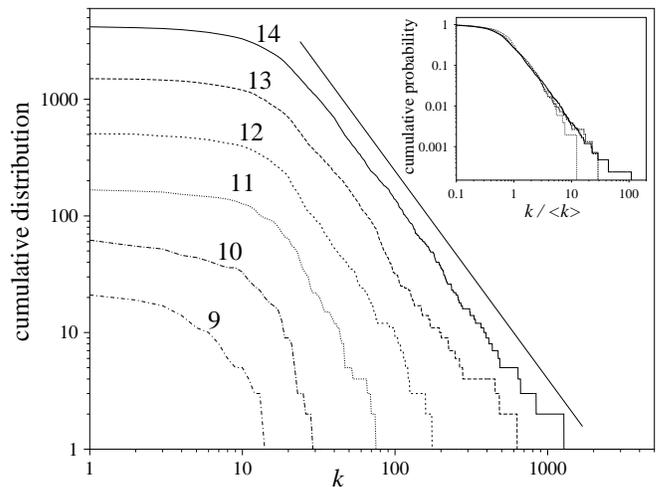}
\caption{
The cumulative distribution for the number of nodes that have more
than $k$ connections. The curves correspond to clusters of different sizes, as labelled.
An additional straight line with slope $-(\gamma-1)$, where $\gamma$=2.78, has
been plotted to emphasise the power law tail.
In the inset the cumulative probability distribution for the 12-, 13- and 14-atom clusters
is plotted against $k$ normalized by its average value, $\langle k \rangle$, to 
bring out the universal form of the distribution.
}
\label{fig:degree}
\end{figure}

If we now examine the distributions for the numbers of connections for each node
we find that as the size of the cluster increases a clear power-law tail develops, which
has a universal form independent of the cluster size (Fig.\ \ref{fig:degree}). 
The exponent of the power law, $\gamma$=2.78, is similar to other scale-free networks.\cite{Albert01}
The cause of the random-graph like scaling of the average separation is thus
the scale-free topology of the ISN.
The network is extremely heterogeneous with a few hubs that have
a very large number of connections, but with the majority of nodes only connected to 
a relatively small number of other minima.

This is a particularly surprising result 
because all other scale-free networks are dynamic in origin. 
They grow and change over time, be it on an almost continuous basis as in the WWW 
or on evolutionary time scales in the case of biochemical networks. 
Even the recently introduced deterministic scale-free networks are based on an
iterative growth procedure.\cite{Barabasi01,Dorogovtsev02}
Furthermore, models of network growth\cite{Barabasi99,Albert01} 
and studies on the time evolution of real networks\cite{Newman01d,Jeong01b} strongly suggest that
the heterogeneity at the heart of the scale-free topology develops as 
a result of new nodes preferentially linking to those nodes which 
have many connections, be they much-cited papers or popular web-sites.
However, the network associated with a potential energy landscape is static. It is simply
determined by the form of the interatomic interactions and the number of atoms in the cluster.

\begin{figure}
\includegraphics[width=8.6cm]{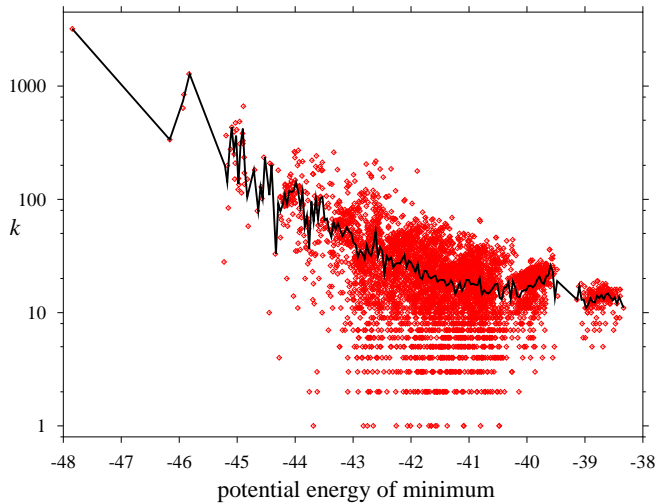}
\caption{
The dependence of the degree of a node on the potential energy 
of the corresponding minimum for the 14-atom cluster. The data points are 
for each individual minimum and the solid line is a binned average.
}
\label{fig:LJ14}
\end{figure}

The source of the heterogeneity in the ISNs is apparent from Fig.\ \ref{fig:LJ14},
where we see that the degree of a node increases somewhat faster than exponentially 
as the energy of the minimum decreases.\cite{mindist} 
The low-energy minima act as the hubs in the network. 
Thus, for the 14-atom cluster, 76\% of the nodes in the network are connected to the global minimum.
To measure the extent of the catchment basin around a minimum,
we can calculate the distance in configuration space to all the 
transition states connected to a minimum.
For LJ$_{14}$ we find that this distance is 2.7 times larger for the global minimum than
for the surrounding minima. 
When the multi-dimensionality of configuration space is taken into account this result
suggests that the hyperarea of this catchment basin is many orders of magnitude 
larger than the average.
Similarly, it has been previously found that on average the basin area falls off 
approximately exponentially with the energy of a minimum.\cite{Doye98e}

These results show that the global topology and topography of the potential energy landscape 
are intimately connected because the deep minima have very large catchment basins which are 
connected to lots of smaller basins around their edges. 
By contrast, if a potential energy landscape were flat and all basins of 
attraction had the same area the scale-free topology would be lost.
For example, an investigation of the network topology of the configuration space 
of a {\em non-interacting} lattice polymer\cite{Scala01b} found 
the connectivity distribution to be a Gaussian.\cite{Amaral00} 

This contrasting example naturally leads one to ask how general is the topology 
that we have found for the Lennard-Jones clusters.
Although such a question can only be definitively answered by similar analyses for a 
variety of systems, there is nothing ``special'' about the Lennard-Jones potential
and so there is no reason why similar behaviour should not be seen for other 
materials, as long as there are no constraints present that would
prevent the formation of the high degrees associated with the hubs.  
A polymer provides an example of the latter, because the connectivity of the chain 
limits the number of transition states that can surround a minimum. 
For example, a similar analysis for Lennard-Jones polymers\cite{Calvo02a} 
did not find a power-law tail to the degree distribution 
(although it was still longer than exponential).\cite{JDunpub}
There is no equivalent of many of the transition states for the equivalent Lennard-Jones 
cluster because they involve the breaking of the polymer chain.

The scale-free topology of the ISN is 
potentially good news for global optimization, the task of locating the global minimum.
Even though the number of minima increases exponentially with the size of the system,\cite{Still99}
the average number of steps in the shortest path to the global minimum grows 
sublinearly with system size. Of course, finding this path is not necessarily easy.
Our calculations of the shortest paths required information on the global structure of the 
potential energy landscape, whereas a global optimization algorithm usually takes a
step based on only local information.

Some path finding strategies to efficiently navigate scale-free networks have been suggested
that make use of the fact that most of the shortest paths pass through the highly-connected
hubs.\cite{Adamic01,Kim02} In our case the clear link between the topology and 
topography of the potential energy landscape provides an additional advantageous strategy. 
A downhill step to a lower-energy minimum is likely to take 
one to a minimum that is more connected and closer to the global minimum.
How well-obeyed the latter correlation is, depends upon the global topography
of the potential energy landscape and is a good indicator of the difficulty
of global optimization. Thus, when the landscape is like a single funnel, global 
optimization algorithms can achieve near to the ideal scaling.
For example, the basin-hopping algorithm can locate the global minimum 
of the 55-atom Lennard-Jones cluster after on average less than 150 minimizations 
when started from a random configuration, 
even though there are an estimated $10^{21}$ minima.\cite{Doye98e} 
The increasing number of links as the energy decreases evident in Fig.\ \ref{fig:LJ14}
further adds to the efficacy of a funnel in guiding the system towards its bottom,
and provides clear evidence of the convergence of pathways into the hub at 
the funnel bottom that is often postulated.
By contrast, when an energy landscape has multiple funnels 
and there is a tendency to enter a funnel that leads the system to a low-energy minimum
that is far from the global minimum, global optimization can be very difficult.

The topology of the ISN will of course significantly affect the dynamics. 
This connection can be probed for very small systems where the network can be completely
characterized and the inherent structure dynamics obtained by a master equation approach.
However, this approach is not practical for the system sizes that are of most interest. 
Therefore, models of protein folding and the glass transition usually have to assume a 
simplified topology for the interstate dynamics,\cite{Shak89,Kohen00} 
or relate the dynamics to static quantities through phenomenological equations, 
such as the Adam-Gibbs equation which relates the relaxation time in supercooled liquids 
to the configurational entropy.\cite{AdamG65}
To fully unlock the potential insights from the inherent structure view of the dynamics, a means
of statistically modelling the network topology from a partial characterization of 
the potential energy landscape is therefore needed.
Our results could significantly advance this goal.

The author is grateful to Emmanuel College, Cambridge 
and the Royal Society for financial support.


\begin{thebibliography}{10}
\bibitem{cluster}
For the smaller clusters the clustering coefficients are not that much larger
  than random simply because these small networks are so highly connected.
\bibitem{mindist}
This dependence of the degree on the energy of a minimum combines with an
  energetic distribution of minima, which is roughly Gaussian at the most
  common energies\cite{Sciortino99a,Buchner99a} but with a low-energy tail that
  falls off less rapidly, to give the power law tail of the degree
  distribution.
\end{thebibliography}
\end{document}